\documentstyle[emulateapj,apjfonts]{article}

\def\ltsim{ \,{}^<_\sim\, }
\def\gtsim{ \,{}^>_\sim\, }
\def\etal{et~al.}

\slugcomment{accepted for publication in the Astrophysical Journal}
\lefthead{Kavelaars \etal }
\righthead{Hubble Constant from GCLF}

\begin{document}
\large

\title{The Globular Cluster Systems in the Coma Ellipticals. I: 
The Luminosity Function in NGC 4874, and Implications for Hubble's Constant\altaffilmark{1}}

\altaffiltext{1}{Based on observations 
obtained with the NASA/ESA {\it
Hubble Space Telescope}, obtained at the Space Telescope Science
Institute, which is operated by the Association of Universities for
Research in Astronomy, Inc., under NASA contract NAS 5-26555.}

\author{J.~J.~Kavelaars and William E.~Harris}
\affil{Department of Physics and Astronomy, McMaster University, 
Hamilton ON L8S 4M1, CANADA;  kavelaar,harris@physics.mcmaster.ca}

\author{David A. Hanes}
\affil{Department of Physics, Queen's University, Kingston ON K7L 3N6, CANADA; hanes@astro.queensu.ca}

\author{James E. Hesser}
\affil{Dominion Astrophysical Observatory, Herzberg Institute of Astrophysics, National Research Council, 5071 West Saanich Road, Victoria BC V8W 3P6, CANADA; jim.hesser@hia.nrc.ca}

\and

\author{Christopher J. Pritchet}
\affil{Department of Physics and Astronomy, University of Victoria, Box 3055, Victoria, BC V8W 3P6, CANADA; pritchet@phys.uvic.ca}

\begin{abstract}
We have used deep HST/WFPC2 images in $V$ (F606W) and
$I$ (F814W) to measure the luminosity distribution of the globular clusters
in NGC 4874, the central cD galaxy of the Coma cluster.  We find 
the ``turnover'' point of the globular cluster luminosity function 
(GCLF) to lie at $V = 27.88 \pm 0.12$, while the overall GCLF 
shape matches the standard
Gaussian-like form with dispersion $\sigma_V = 1.49 \pm 0.12$.
We use the GCLF as a standard candle by matching the turnover points in
NGC 4874 and another Coma elliptical, IC 4051, with
those of the giant ellipticals in the Virgo cluster (M87 and five others).
The result is $\Delta(m-M)$(Coma - Virgo) $= 4.06 \pm 0.11$ magnitudes,
which converts to a Coma distance $d = 102$ Mpc if the Virgo 
distance modulus is $(m-M)_0 = 30.99 \pm 0.04$.
The Hubble constant which emerges from our GCLF measurement is then
$H_0 = (69 \pm 9)$ km s$^{-1}$ Mpc$^{-1}$.  We confirm this $H_0$ value
with a novel presentation of the ``Hubble diagram'' for GCLFs in giant
E galaxies.  Measurements of additional
GCLFs in the Coma ellipticals, as well as calibrating galaxies in
Virgo and Fornax, have excellent potential to 
refine this result in the near future. 
\end{abstract}


\section{Introduction}

One of the simplest and longest-range stellar standard candles for distance
measurement is the globular cluster luminosity function (GCLF).  The
number distribution of globular clusters per unit magnitude has long
been known to have a unimodal and roughly symmetric form, with a 
peak frequency or ``turnover'' in the range $M_V^0 \sim -7.4 \pm 0.2$
in both spiral and elliptical galaxies.
Theoretical concerns have often been expressed that 
the mass distribution of globular clusters (and thus the GCLF) 
might be expected to 
depend on external factors such as galaxy type, 
metallicity, or radial location within a galaxy; but the empirical
evidence strongly suggests that these factors exert
remarkably little influence on the peak of the GCLF 
(see \cite{jac92}; \cite{whi96}; \cite{ash95}; \cite{har99} for
thorough discussions).
With the imaging capabilities of the Hubble Space
Telescope, the GCLF turnover is within reach for any galaxy up to at
least 120 Mpc, beyond the regime of local peculiar motions 
which might strongly bias measurements of the Hubble constant $H_0$.

For purposes of remote distance calibration and estimation of
$H_0$, the globular cluster populations in
giant elliptical galaxies are by far the most interesting targets,
simply because of the sheer size of their cluster populations and thus
the ability to define the GCLF shape and turnover with high statistical
confidence.  It is also a fortunate coincidence for this distance-scale
method that the very most populous globular cluster systems 
inhabit the supergiant ellipticals that lie at the centers of rich
clusters of galaxies, which are the very objects that 
are the main landmarks in the Hubble flow.  

In this paper we present our new measurement 
of the GCLF turnover for the globular cluster system in NGC 4874, the
central cD elliptical in the Coma cluster, and use it to estimate
$H_0$.  The bright end of the GCLF in NGC 4874 has already 
been studied with ground-based imaging, which indicated that it
indeed has a large globular cluster system (\cite{har87}; \cite{tho87};
\cite{bla95}).
With the much deeper HST photometric limits, we could therefore
confidently expect to garner a huge population of clusters to define the
GCLF.

\section{The Data}

Our raw dataset consists of 18 $V$ (F606W) and 10 $I$ (F814W) images of
various exposure lengths (see Table~\ref{tab:obs}), taken 1997 August 16
and 24 (program GO-5905) with the WFPC2 camera.The long exposures were sub-pixel-shifted 
(dithered) in a pentagonal pattern
by fractional pixel amounts in order to reconstruct clean
composite images free from cosmic-ray contamination and
bad-pixel artifacts.  The $V$ exposures, totaling 20940 sec, were taken
to probe the GCLF deeply enough to resolve the turnover point, while the
$I$ exposures, totaling 8720 sec, were used to 
define the color (metallicity)
distribution for the brighter end of the cluster system. The color
distribution, spatial structure of the GCS, and the specific frequency
will be discussed in Paper II (\cite{har99a}).  Here, we analyze the
GCLF and use it to estimate the distance to Coma and thus $H_0$.

\begin{deluxetable}{lllc}
\tablewidth{0pt}
\tablecaption{\label{tab:obs}{\it HST} Observing Log for GO 5905 }
\tablehead{
\colhead{R.A.} & \colhead{Decl.} & & \colhead{Exposure Times} \\
\colhead{(J2000)} & \colhead{(J2000)} & \colhead{Filter} &
\colhead{(s)}
}
\startdata
12 59 33.43 & +27 57 43.3 & F606W & $3\times 180$ \\
            &             &       & $2\times 1100$ \\ 
            &             &       & $14\times 1300  $ \\
& & F814W & $4\times 230 $ \\
& &       & $6 \times 1300 $ \\
\enddata
\end{deluxetable}

To maximize the total globular cluster population falling within the
WFPC2 field of view, we placed the center of the PC1 CCD on the nucleus
of NGC 4874.  A few other large elliptical galaxies projected near the
center of Coma fall on the outskirts of the WF2,3,4 CCDs; however,
these proved not to have significant numbers of globular clusters of their
own and thus did not contaminate the NGC 4874 sample.  Small areas 
surrounding them were, in any case, masked out in all
subsequent data analysis.

We first retrieved the raw data from the HST archive located at the
CADC\footnote{Canadian Astronomy Data Centre, operated by the
Herzberg Institute of Astrophysics, National Research Council of
Canada.}.  The CADC pipeline preprocessed the
images at this point, with the best calibration images then 
available. We then combined the exposures in pairs 
(the pairs of long exposures within each orbit) to define a first set 
of frames reasonably free of cosmic-ray contamination.

The vast majority of detected objects on the frames are the globular
clusters in the halo of NGC 4874.  At the $\sim 100-$Mpc distance of Coma 
they appear as unresolved point sources even in the PC1 frames, so
it is readily possible to perform conventional point-spread function 
(PSF) photometry on the frames.  We constructed an independent PSF
for each image and each of the four WFPC2 CCDs.  
With DAOPHOT and ALLSTAR (\cite{ste94}) we then generated separate 
lists of candidate starlike (that is, unresolved) objects on each frame.
These coordinate
lists were used to determine the (small) offsets and rotations 
to map the images onto a common re-registered coordinate system.
Next, Stetson's (1994) MONTAGE2 code was used to define a
``master'' image in each filter as the median (i.e., 50th percentile) of
the individual exposures.  Finally, we 
summed the master $V$ and $I$ images 
to generate a single deep, contamination-free image 
which maximized the available flux for object detection.

\begin{figure*}
\epsscale{0.5}
\plotone{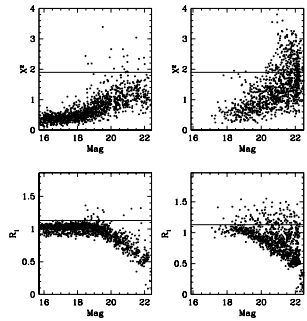}
\figcaption[f1.eps]{Image moments $\chi$ and $r_1$ plotted against magnitude.
The two panels on the left show artificial-star sequences, while the two
on the right show the real objects measured on the WFPC2 frames.
Filled circles are objects assumed to be ``starlike'', while the open
circles are ``nonstellar'' and culled from the sample.
\label{fig:image_moments}}
\end{figure*}

The photometry must also be designed to
avoid false detections and nonstellar objects (noise spikes, small, faint
background galaxies, or even compact dwarf galaxies within Coma itself).
The real and artificial-star measurements were therefore combined with
image shape parameters (the DAOPHOT parameters SHARP and 
$\chi$, and the radial image moment $r_1$ defined by \cite{kro80} and 
\cite{har91}).
Several numerical trials were carried out with the artificial-star data
to vary the detection threshold (in units of $\sigma_s$, the RMS 
scatter of the sky background) and the shape selection parameters,
and thus to determine the highest values of these thresholds which would not
deteriorate the detection efficiency of genuine starlike objects (see
Figure \ref{fig:image_moments}). 
 We adopted a DAOPHOT/FIND detection threshold of 
$3.5 \sigma_s$ above sky (c.f. \cite{ste87}), a level which recovered virtually
all the brighter input artificial stars while adding almost no false 
detections from noise.  
The adopted boundaries for $\chi$, SHARP, and $r_1$ resulted in
the culling of about 5\% of the artificial stars, mostly at
the faintest levels where the distinction between starlike and
nonstellar objects by image shape classification also becomes difficult.
At the faint end as well, some unwanted nonstellar objects end up
being scattered into the ``starlike'' category, but these must be
statistically removed from the final GCLF by subtraction of the
background luminosity function (see below).

With the master list of starlike objects now determined, we used
the ALLFRAME code (\cite{ste94}) to measure these objects 
on the original set of F606W and F814W
images, employing the individual PSFs for each frame as determined
previously.  To transform the averaged instrumental
magnitudes to the standard $VI_{KC}$ system we employed
the normal transformation 
equations in Holtzman \etal\ (1995)\nocite{hol95}.  
For the faint objects lacking 
any $I-$band measurement, we assumed a $(V-I)$
colour of 0.95, the observed mean color of the brighter clusters (see
below).  The basic transformation used was
\begin{eqnarray}
V &  = &  m(F606W) + 0.254 (V-I) \nonumber \\ 
  &    & \mbox{}  + 0.012 (V-I)^2 + 22.098   \nonumber\\
I & =  & m(F814W) - 0.062 (V-I) \nonumber \\ 
  &    & \mbox{} + 0.025 (V-I)^2 + 20.839, \nonumber 
\end{eqnarray}
\noindent to which we then added the individual zero-point 
terms for each chip as listed in Table~\ref{tab:trans}.

\begin{deluxetable}{lllll}
\tablecaption{\label{tab:trans}Transformation terms for WFPC2 CCDs}
\tablehead{
\colhead{} & \colhead{PC1} & \colhead{WF2} & \colhead{WF3} & \colhead{WF4}
}
\startdata
Geometric Factor & -0.0071 & -0.0005 & 0 & -0.0014 \\
Gain Factor & 0.7455 & 0.7542 & 0.7558 & 0.7279 \\
Aperture Correction \\
~~F606W & $-0.26\pm0.10$ & $-0.28\pm0.05$ \\  
~~F814W & $-0.25\pm0.07$ & $-0.28\pm0.05$ \\ 
\enddata
\end{deluxetable}

In these equations, we have added the standard value of 0.05 mag
to the published zero-points to account for the deferred charge transfer
effect that produces an offset between long and short exposures
(e.g., Stetson \etal\ (1999)\nocite{ste99}).
As a direct check on this offset, we intercompared the magnitudes of several
of the brightest stars on our long and short exposures, and found
$$ (m_{1300s} - m_{180s})_{F606W} = -0.06\pm0.02, $$
$$ (m_{1300s} - m_{230s})_{F814W}  = -0.06\pm0.03.$$ 
That is to say, magnitudes
determined using short exposures (like those used for 
the calibration of \cite{hol95}) 
tend to be fainter than those determined using long exposures; our
internal tests confirm the $\simeq 0.05-$mag offset normally used.

\section{The GCLF}

Our final catalog of objects with deep $V-$band measurements is still
contaminated to some extent by a few foreground stars (nearly
negligible) and by faint, extremely small galaxies that might have
crept through the image classification procedure described above.
To define a ``background'' population, we simply use those objects which
lie on the radial outskirts of our field, more than 
75 arcseconds (corresponding roughly to 40 kpc) from the center of NGC 4874. 
At these large radii, we find that the number density of globular
clusters is still declining; that is, 
the total extent of the GCS evidently spills
well beyond the borders of our single WFPC2 field.  However, the object
density within the background area is only a small fraction (14\%) of the
density in the core region (the area covered by the PC1 chip), so 
we can use the outer corners for background knowing that it will produce
only a small over-subtraction of the globular cluster population itself.

\begin{figure*}
\epsscale{0.5}
\plotone{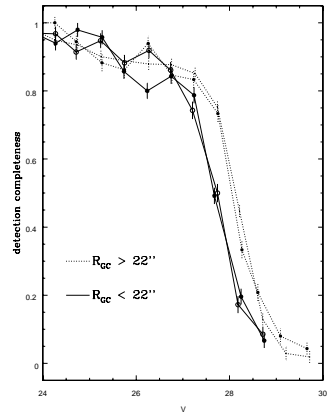}\caption[f2.eps]{Detection efficiency as determined by artificial star tests.
The inner rings (mostly drawn from the PC chip) displays brighter
cutoffs, which are caused by the higher halo light level at small radii.
\label{fig:recf}}
\end{figure*}

This raw GCLF must be corrected for detection incompleteness at the 
faint end.  The detection probability is, in turn, dependent on the
local level of background light and thus is a function of 
the distance from the galaxy center.  Figure
~\ref{fig:recf} shows the completeness fraction for four radial zones
($3\arcsec\le r < 13\arcsec, 13\arcsec\le r <  22\arcsec, 22\arcsec\le r < 50\arcsec, 
50\arcsec\le r < 100\arcsec$).
Clearly, the inner zones -- mostly within PC1 and within the bright
galaxy envelope -- have noticeably brighter, but consistent, completeness 
cutoff levels than in the outer annuli.  This radial dependence 
was explicitly folded in to the completeness corrections to the raw
data.

\begin{figure*}
\epsscale{0.5}
\plottwo{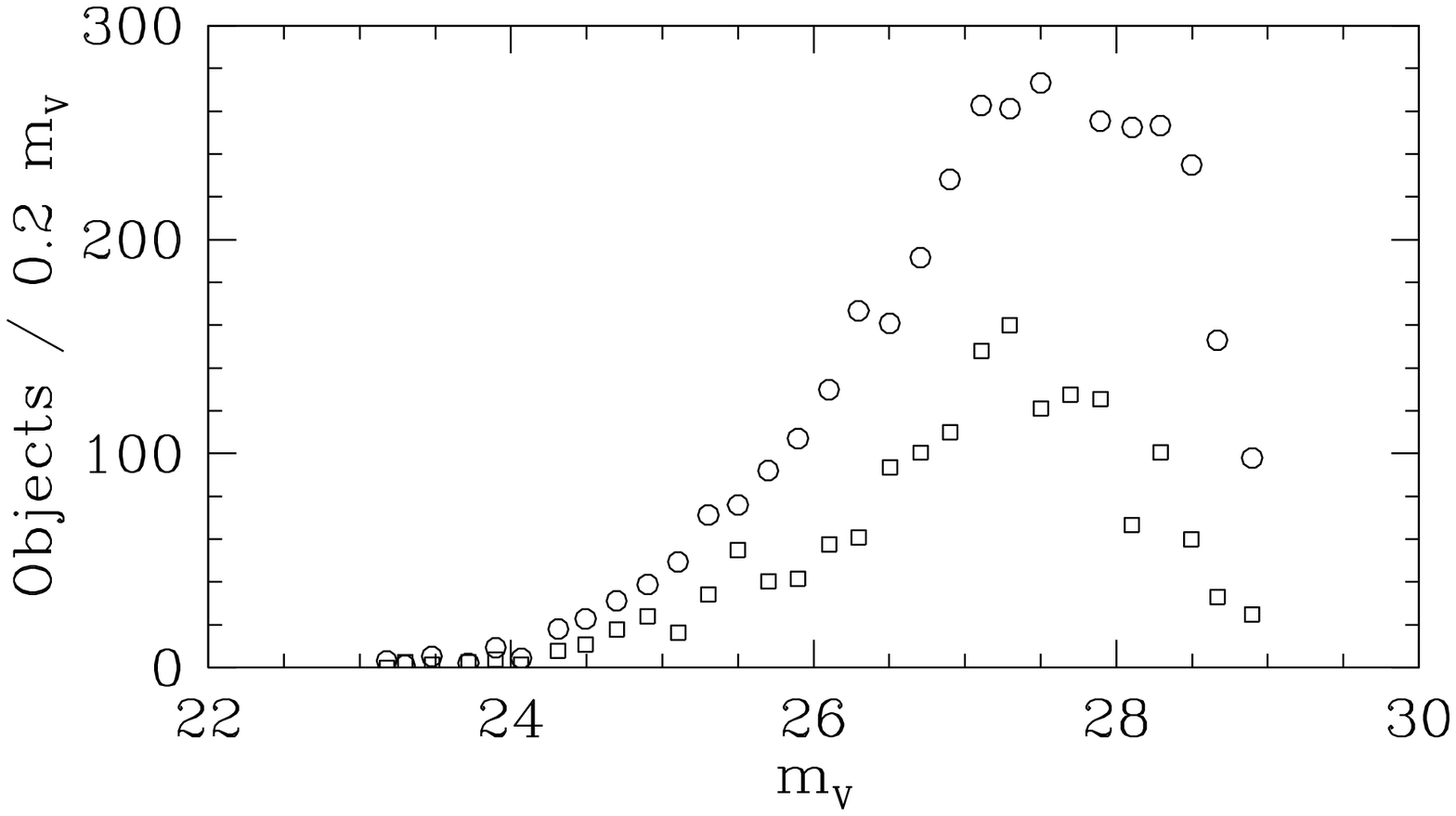}{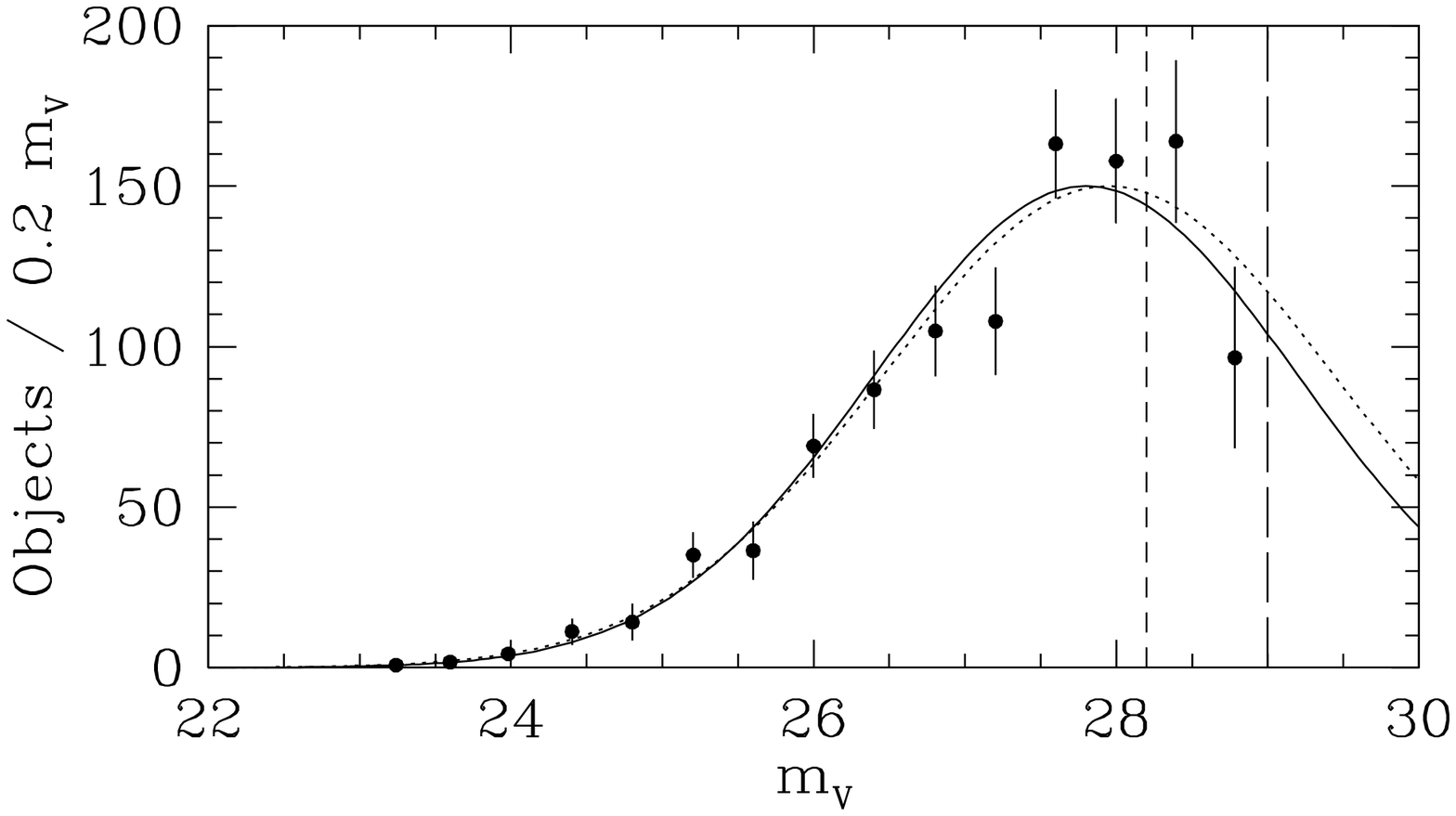}
\caption[f3a.eps,f3b.eps]{
{\it Left panel:}  The data defining the globular cluster 
luminosity function (GCLF).  The open circles indicate the 
luminosity function of all starlike objects within $75\arcsec$ 
of the galaxy center in $0.2-$magnitude bins and corrected for
incompleteness, while
the open squares represent the LF at radii larger than $75\arcsec$ with the 
counts normalized to the area of the inner region.
{\it Right panel:}  The points show the 
the background-subtracted and completeness-corrected 
GCLF used for the $\chi^2$ fits, now binned in $0.4-$magnitude steps.
The solid line is a Gaussian of width
$\sigma_V = 1.4$ and peak $V^0 = 27.8$ (the ``constrained'' fit described
in the text), while the dotted line is a Gaussian with $\sigma_V = 1.49$ and
peak $V^0 = 28.0$ (the unconstrained maximum-likelihood fit).
The short-dashed line indicates the 50\% completeness level while the
long-dashed line is the 5\% completeness level.
\label{fig:gclf_binned}}
\end{figure*}

The full completeness-corrected and background-subtracted luminosity
functions are shown in Figure~\ref{fig:gclf_binned} and Table~\ref{tab:gclf_data}.  The limit
of our deep $V$ photometry reaches the turnover point (GCLF peak)
or just past it.

\begin{deluxetable}{lllll}
\tablewidth{0pt}
\tablecaption{\label{tab:gclf_data}
Luminosity function data}
\tablehead{\colhead{$\langle m^0_V\rangle $} & \colhead{Inner} & \colhead{$\pm $} & \colhead{Outer} & \colhead{$\pm $ } } 

\startdata
     23.2   &   3    &   2    &       0  &   0 \\
     23.3   &   1    &   1    &       2 &   1 \\
     23.5   &   5    &   2    &       1 &   1 \\
     23.7   &   2    &   1    &       2 &   1\\
     23.9   &   9    &   3    &       3 &   2\\
     24.1   &   4    &   2    &       1 &   1\\
     24.3   &   18   &   4    &       7 &   3\\
     24.5   &   22   &   5    &      10 &   3\\
     24.7   &   31   &   6    &      17 &   4\\
     24.9   &   38   &   6    &      23 &   5\\
     25.1   &   49   &   7    &      16 &   4\\
     25.3   &   71   &   9    &      34 &   6\\
     25.5   &   76   &   9    &      55 &   9\\
     25.7   &   92   &   10   &      40 &   7\\
     25.9   &   107  &   11   &      41 &   7\\
     26.1   &   129  &   12   &      57 &   8\\
     26.3   &   166  &   14   &      60 &   9\\
     26.5   &   160  &   13   &      93 &  11\\
     26.7   &   191  &   15   &     100 &  12\\
     26.9   &   228  &   16   &     109 &  12\\
     27.1   &   262  &   18   &     147 &  14\\
     27.3   &   261  &   17   &     160 &  16\\
     27.5   &   273  &   18   &     120 &  12\\
     27.7   &   301  &   21   &     127 &  14\\
     27.9   &   255  &   21   &     125 &  15\\
     28.1   &   252  &   25   &      66 &  12\\
     28.3   &   253  &   28   &     100 &  17\\
     28.5   &   235  &   34   &      59 &  17\\
     28.7   &   153  &   36   &      32 &  14\\
     28.9   &   98   &   37   &      24 &  17\\
\enddata
\end{deluxetable}

The standard candle of the GCLF method is the luminosity level of the
turnover or peak point, which we call $V^0$.
The immediate goal for this discussion is therefore to estimate
this point as accurately as possible given that we see only the
bright half of the GCLF and the turnover region itself, and not the
faint half.  Conventionally, simple interpolation functions such as
a Gaussian (e.g., \cite{jac92}) or $t_5$ (\cite{sec93}),
both of which have two free parameters ($V^0$ and the dispersion
width), have been used for this purpose since they have repeatedly been
shown to match the region near the turnover quite accurately.

For cases such as ours where the photometric limit
$V(lim)$ is fairly close to $V^0$, it is important to note that
attempts to solve {\it simultaneously} for both the peak $V^0$ and dispersion 
$\sigma_V$ of the interpolation function are problematic, because the two
parameters are correlated (see \cite{sec93}; \cite{han87})
and tend to produce overestimates
of both the turnover and dispersion when the fit is constrained by only one
side of the GCLF.  A systematically more accurate
procedure is to {\it adopt} a value for $\sigma_V$ 
and to solve only for the turnover magnitude.  Such an approach is
expected to work essentially because, for giant ellipticals,
the GCLF dispersion in its Gaussian form is observed to be 
highly consistent from one galaxy to another.  For 13 gE galaxies with
well measured GCLFs, Whitmore (1996) and Harris (1999) find
$\sigma_V = 1.4$ with an uncertainty of just $\pm 0.05$.
In particular, this dispersion value fits the very thoroughly studied 
Virgo giant M87 quite well (see \cite{har98b}; \cite{kun99}).  A further 
application of Blakeslee \& Tonry's (1995)\nocite{bla95} variant of the SBF 
method to the GCLF may permit the removal of this constraint on $\sigma_V$.

For an initial fit, we use a Gaussian function and 
employ a $\chi^2$ minimization to the binned data (second panel
of Fig.~\ref{fig:gclf_binned}).
Table~\ref{tab:gclf} summarizes the variety of parametric fits
we obtained under the constraints listed.\begin{deluxetable}{lll}
\tablewidth{0pt}
\tablecaption{\label{tab:gclf}Gaussian representations of the GCLF }
\tablehead{
\colhead{$m^0_V$} & \colhead{$\sigma_V$} & \colhead{$\chi^2$} }\
\startdata
$28.08\pm0.3$ & $1.5\pm0.3$ & 1.4 \\
$27.90\pm0.08 $& $1.4$ & 1.4 \\
$28.00^{+0.19}_{-0.22}$ & $1.49\pm0.11$& ML\tablenotemark{b} \\
$27.88\pm0.10$ & 1.4 & ML\tablenotemark{c} 
\tablenotetext{a}{Values given without error ranges were constrained during fitting, error ranges given are 1 $\sigma_p$ uncertainties}
\tablenotetext{b}{Maximum likelihood analysis}
\tablenotetext{c}{Constrained maximum likelihood analysis}
\enddata
\end{deluxetable}
Employing the entire data range, i.e.~including data even beyond 
the 50\% completeness
level, typically results in a best-fit Gaussian which is quite broad
and whose peak is biased towards fainter magnitudes.  This effect 
has been seen many times in the literature and is
the result of the parametric correlation between $V^0$ and $\sigma_V$
mentioned above.  
The $\chi^2$ fit is much more robust if only those objects brighter
than the 50\% completeness limit are used (see row one of the table)
and the width is kept fixed at the canonical value of 
$\sigma_V = 1.4$ (row two of the table).
As seen in Table~\ref{tab:gclf}, this constrained fit results in a
$\chi^2$ which is little different from that of the unconstrained fit.

\begin{figure*}
\epsscale{0.5}
\plotone{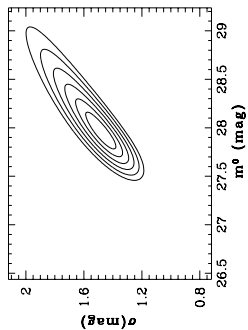}
\caption[f4.eps]{Maximum-likelihood contours (dispersion $\sigma_V$ against
turnover magnitude $V^0$) for a Gaussian representation of 
the GCLF of NGC 4874. The most probable solution is $\sigma_V = 1.49$, $V^0
= 28.00$; the innermost contour is 0.5 $\sigma_p$ (standard deviations) with 
steps of 0.5 $\sigma_p$ between contours.  Note that the two parameters
are correlated, and the contours are asymmetric such that higher values
of each are favored over lower ones.  
\label{fig:ml}}
\end{figure*}

Next we employ 
a maximum-likelihood fit to the raw data following the precepts
of Secker \& Harris (1993)\nocite{sec93}.  Here, the model Gaussian curve is convolved
with the photometric completeness and photometric error functions and
then matched to the raw, uncorrected GCLF.  
The results are shown in Table~\ref{tab:gclf} and Figure~\ref{fig:ml}.

An unconstrained fit for both $V^0$ and $\sigma_V$
(now excluding only those data fainter than the 5\% completeness limit) 
again results in a fainter
turnover and broader dispersion much as did the unconstrained
$\chi^2$ fit to the binned data.  Also given in the table 
is the range of values of
$V^0$ consistent (at the 1 $\sigma$ level) with a Gaussian of
width $\sigma_V = 1.4$. This range of values mimics that of the
constrained $\chi^2$ fit but make use of a larger magnitude range of the data.

The results of the unconstrained fits and those with $\sigma_V$ set to
1.40 appear in the second panel of Figure~\ref{fig:gclf_binned}. They
are very similar.  The history of GCLF analysis can be a guide in our
final choice of the turnover luminosity.  
For data which reach only to the turnover point or just past it, the
unconstrained fits have almost always proven to yield overestimates of
both the turnover and GCLF dispersion, with the benefit of hindsight
and deeper data.  We suggest that the most prudent choice is that of
the constrained fits.  There is little to choose between the
maximum-likelihood and $\chi^2$ approaches; we adopt $V^0 =
27.88\pm0.12$ and $\sigma_V = 1.49^{+0.15}_{-0.10}$ (fitting errors
only), where the range of $\sigma_V$ is that allowed via the maximum
likelihood analysis.

Finally, we have followed exactly the analysis outlined above but
this time using only the data from the inner region and without first
subtracting a background.  Fits to these ``un-subtracted'' data result
in parameters identical to those we report here for the background
subtracted $\chi^2$ fits.  The only difference found is in the allowed
range of $\sigma_V$ as reported via a maximum-likelihood analysis
which reports an allowed range of $1.7 < \sigma_V < 2.0$.  This
extension of $\sigma_V$ to larger values is the expected result if we
assume that the majority of our background is derived from the
mis-classification of faint background sources as stellar, thus
elevating the faint-end of the luminosity function.  By once again
constraining the maximum-likelihood fits of the luminosity function of
the inner region, requiring that $\sigma_V = 1.40$, we recover the
results reported here for peak of the background subtracted
distribution.

The Coma cluster is almost at the North Galactic Pole, with
nearly negligible foreground extinction. 
Here we adopt the same value as used by
Baum \etal\ (1997) in their discussion of the GCLF in IC 4051, namely
$E(B-V) = 0.01$ and $A_V = 0.03\pm0.01$.  We also include the minor $K$
(cosmological redshift) correction of $0.03\pm0.01$ as detailed 
in their analysis.  Thus our turnover magnitude, dereddened and
$K$-corrected, is $V^0 = 27.88 - 0.06 \simeq 27.82\pm0.13$.

Two other Coma ellipticals have GCLFs measured from deep HST photometry:  
NGC 4881 (\cite{bau95}) and IC 4051 (\cite{bau97}; \cite{woo99}).
Both of these are normal, relatively isolated giant ellipticals 
on the outskirts of the cluster core.  
The NGC 4881 data do not reach faint enough to
define the GCLF turnover unambiguously, and we will not use it here.  
However, the IC 4051 data are quite comparable in depth 
to our NGC 4874 measurements. For it, Baum \etal\ (1997) 
found $V^0 = 27.79$
from the population of objects on only the PC1 frame\footnote{We have 
adjusted their result 
to account for the 0.05 magnitude shift between long and
short exposures as noted above; see e.g., Stetson \etal\ (1999).}. Woodworth \& 
Harris (1999), from an independent reduction of the same raw data but
now including all four WFPC2 frames (thus doubling the globular cluster
population compared with the PC1 alone)
determined $V^0 = 27.77\pm0.20$ (their quoted error includes the
uncertainty of the fitted Gaussian parameters).  We adopt for IC 4051 a mean 
value of $V^0 = 27.72\pm0.20$ after the $K$-correction and 
extinction are subtracted.

The measured turnover levels of both galaxies -- NGC 4874 and IC 4051 --
agree at the $0.1-$magnitude level, well within their
combined uncertainty of $\pm0.25$ mag.  A straight mean of
the two gives $V^0$(Coma) $= 27.77\pm0.10$, which we use for the
following discussion.

\section{Calibrating the Distance to Coma}

Once an intrinsic luminosity for $M_V^0$(GCLF) is adopted, the distance
to Coma follows immediately.  In principle, we could take the value for
$M_V^0$ from the Milky Way GCLF and therefore jump from the Milky Way to Coma in
a single step, bypassing the entire chain of distance indicators
which leads through the Local Group, Virgo, Fornax, and so forth (e.g., \cite{bau97}).
However, the fundamental objection to such an approach is simply that we would be
comparing two very different types of galaxies, in which it is possible
that the globular cluster formation and evolution processes have been
different.  To avoid such concerns
as much as possible and place our result on astrophysically more
defensible ground, we choose instead to 
compare only giant ellipticals with giant ellipticals.

In essence then, to calibrate $M_V^0$ we need to have well
determined GCLFs in giant ellipticals (and preferably cD's as well)
whose distances are themselves well established from more fundamental stellar
standard candles.
The largest nearby collections of such gE's are in the Virgo and Fornax
clusters.  Fortunately both of these have 
central cD-type galaxies (M87 in Virgo, NGC 1399 in Fornax)
as well as many other giant ellipticals which provide highly effective
tests of the empirical galaxy-to-galaxy scatter of the GCLF method 
(\cite{whi96}; \cite{har99}).  

Distance indicators applicable to these nearby galaxies are the red-giant-branch 
tip luminosity (TRGB), surface brightness
fluctuations (SBF), planetary nebulae (PNLF), and of course Cepheid variables.
These four methods have the strongest observational claims to 
precisions approaching $\pm 0.1$ mag, and their foundations 
are well understood from basic stellar physics (\cite{jac92};
\cite{lee93}; \cite{fer99a}).  
Results taken from the current literature are summarized
in Tables \ref{tab:virgo_distance} and \ref{tab:fornax_distance}.

\begin{deluxetable}{lll}
\tablewidth{0pt}
\tablecaption{Estimates of distance to Virgo cluster}
\tablehead{\colhead{Technique} & \colhead{($m-M)_0$} & \colhead{Sources}\tablenotemark{a} } 
\startdata
Cepheids & $31.02 \pm 0.04$ & 1,2,3,4,5,6 \\
SBF & $31.02 \pm 0.05$ & 7,8,9,10 \\
PNLF & $30.88 \pm 0.05$ & 11,12 \\
TRGB & $30.98 \pm 0.18$ & 13 \\
& & \\
$\langle m-M \rangle_0$ & $30.99\pm0.03$ \\
\tablenotetext{a}{Sources:  (1) Ferrarese \etal\ 1996
(2) Pierce \etal\ 1994
(3) Saha \etal\ 1996a
(4) Saha \etal\ 1996b
(5) Graham \etal\ 1999
(6) Ferrarese \etal\ 1999b
(7) Tonry \etal\ 1997
(8) Neilsen \etal\ 1997
(9) Pahre \& Mould 1994
(10) Morris \& Shanks 1998
(11) Jacoby \etal\ 1990
(12) Ciardullo \etal\ 1998
(13) Harris \etal\ 1998a }
\label{tab:virgo_distance}
\enddata
\tablecomments{Uncertainties listed are the 1 $\sigma$ scatter among 
referenced sources, or the internal uncertainties quoted by 
the original authors.}
\end{deluxetable}
\begin{deluxetable}{lll}
\tablewidth{0pt}
\tablecaption{Estimates of distance to Fornax cluster}
\tablehead{\colhead{Technique} & \colhead{$(m-M)_0$} & \colhead{Sources}\tablenotemark{a}} 
\startdata
Cepheids & $31.54 \pm 0.14$ & 1 \\
SBF & $31.23 \pm 0.06$ & 2 \\
PNLF & $31.20 \pm 0.07$ & 1,3 \\

& & \\
$\langle m-M \rangle_0$ & $31.30\pm0.04$ \\
\tablenotetext{a}{Sources:  (1) Ferrarese \etal\ 1999b
(2) Tonry \etal\ 1997
(3) McMillan \etal\ 1993 }
\label{tab:fornax_distance}
\enddata
\end{deluxetable}

Intercomparisons among these methods are still best
done within the Virgo cluster, which has the richest variety of
published results.  For the seven published Virgo Cepheid galaxies, 
we reject only the result for NGC 4639 (which at $(m-M)_0 = 31.8$ 
is $\gtsim 0.7$ mag more distant than the average of the other six,
which are in close agreement; see the compilation of \cite{fer99a}.
We also adopt the revised modulus for NGC 4535 as listed by
\cite{fer99b}, containing the small correction for faint-end
incompleteness.) 

As can be seen in
Table~\ref{tab:virgo_distance}, there is remarkable consistency among
the four methods, and we adopt a weighted-average
true distance modulus of $\mu_0$(Virgo) $ = 30.99\pm0.03$. 
We note that this mean
would change to $\simeq 31.01$ if we had adopted the $\sim 0.1-$mag
larger SBF values listed by Ferrarese \etal\ (1999a).  The
quoted uncertainty represents the scatter among the individual
methods and not the (much larger) external uncertainty in
the fundamental distance scale calibration.  
These moduli are based on an adopted distance modulus
$\mu_0 = 18.5 \pm 0.1$ for the LMC, which provides the main underpinning
for the Cepheid P-L relation.  Thorough reviews of the basis for the LMC
modulus, including re-assessments of the 
{\it Hipparcos} parallax data, are given by Carreta \etal\ (1999)\nocite{car99}, 
Harris (1999)\nocite{har99}, and Fernley \etal\ (1998)\nocite{fer98} among others 
and need not be repeated
here.  In light of these discussions, we feel -- perhaps optimistically
-- that the true
external uncertainty in the Virgo distance is close to $\pm 0.2$ mag.

The distance to Fornax is not as well determined:
fewer galaxies have been studied; the differences among the PNLF, 
Cepheid and SBF distance estimates are larger than 
their internal error margins; and a TRGB calibration is not
yet in hand.  Of the three published Cepheid galaxies (again, with
the incompleteness-corrected values from \cite{fer99b}), two
(NGC 1326A, 1365) have published moduli $\sim 0.2$ mag larger than the
SBF or PNLF values, while the third (NGC 1425) is $\sim 0.6$ mag larger.
This discrepancy reinforces the serious concern that the
outlying Cepheid spirals may either be much more widely spread
than the central ellipticals which are most relevant for GCLF studies, 
or at a different mean distance.\footnote{Ferrarese \etal\ (1999a) adopt
a recalibration of the SBF method which gives them a 
Fornax SBF modulus $\sim 0.35$ mag larger than from Tonry \etal\ (1997),
nominally bringing it into better agreement with the mean
of the three Cepheid spirals.  Considerable further work on the Fornax
ellipticals through a variety of techniques is clearly called for to
sort out these large uncertainties.}

In principle, a better way to approach the comparison is
to use methods relating to the ellipticals alone.
The measured GCLF turnovers of Virgo and Fornax E galaxies 
can be used to compute a 
relative Fornax-Virgo distance, with the {\it assumption}
that they have fundamentally the same luminosity. 
Six ellipticals in each cluster have GCLFs measured
to $\gtsim 1$ mag beyond the turnover point,
with results as summarized in Table \ref{tab:vlist}.\begin{deluxetable}{llll}
\tablewidth{0pt}
\tablecaption{GCLF turnover magnitudes for Virgo and Fornax ellipticals}
\tablehead{\colhead{} & \colhead{Galaxy} & \colhead{$V^0$(turnover)} & \colhead{Sources}\tablenotemark{a}} 
\startdata
Virgo & N4472 & $23.87 \pm 0.07$ &  1,2,3 \\
 & N4478 & $23.82 \pm 0.38$ &  4 \\
 & N4486 & $23.67 \pm 0.04$ &  5,6,7,8,9 \\
 & N4552 & $23.70 \pm 0.30$ &  2 \\
 & N4649 & $23.66 \pm 0.10$ &  1 \\
 & N4697 & $23.50 \pm 0.20$ &  10 \\
& & \\
 & Weighted Mean & $23.73 \pm 0.03$ \\

& & \\
Fornax & N1344 & $23.80 \pm 0.25$ & 11 \\
       & N1374 & $23.52 \pm 0.14$ & 12 \\
       & N1379 & $23.92 \pm 0.20$ & 12,13 \\
       & N1399 & $23.86 \pm 0.06$ & 11,12,14,15 \\
       & N1404 & $23.94 \pm 0.08$ & 11,15,16 \\
       & N1427 & $23.78 \pm 0.21$ & 12 \\
& & \\
 & Weighted Mean & $23.85 \pm 0.04$ \\
\tablenotetext{a}{Sources:  (1) Secker \& Harris 1993
(2) Ajhar \etal\ 1994
(3) Lee \etal\ 1998
(4) Neilsen \etal\ 1997
(5) Harris \etal\ 1991
(6) McLaughlin \etal\ 1994
(7) Whitmore \etal\ 1995
(8) Harris \etal\ 1998b
(9) Kundu \etal\ 1999
(10) Kavelaars \& Gladman 1998 
(11) Blakeslee \& Tonry 1996
(12) Kohle \etal\ 1996
(13) Elson \etal\ 1998
(14) Bridges \etal\ 1991
(15) Grillmair \etal\ 1999
(16) Richtler \etal\ 1992 }
\label{tab:vlist}
\enddata
\end{deluxetable}

The difference between the weighted averages of $V^0$ gives
$\Delta \mu$(Fornax-Virgo) $= 0.14 \pm 0.05$.  Adding this
distance offset to the adopted Virgo modulus then
suggests $\mu_0$(Fornax,GCLF) $=31.13\pm0.08$.  This value, in turn,
is in entirely reasonable agreement
with the SBF ($31.23\pm0.06$) and PNLF ($31.20\pm0.14$) determinations of
the Fornax distance.  However, it disagrees strongly with
the mean Cepheid distance or with Ferrarese \etal's recalibrated SBF
value.\footnote{Ferrarese \etal\ (1999a) conclude that the GCLF turnover
luminosity is ``a full 0.5 magnitude brighter'' in Fornax than in Virgo.
This conclusion is a direct result of their high adopted Fornax
distance modulus $(m-M)_0 \simeq 31.6$.  
It should be emphasized that, although they use a less complete GCLF
database than ours, their adopted mean {\it apparent} turnover magnitudes for
both Virgo and Fornax are $\ltsim 0.05$ mag different
from our adopted means in Table \ref{tab:vlist} and thus not the source
of the discrepancy.  
In our view, the best distance modulus to apply to the inner,
GCLF-bearing {\it ellipticals} in Fornax is still very much less certain
than for Virgo. Our contention is that the turnover luminosity difference
between the Virgo and Fornax ellipticals is at the level of 0.2 mag or
less (see the discussion below).}
For the purposes of this discussion, we will defer further use of
the Fornax system and employ only the Virgo ellipticals as our 
calibrators for the GCLF turnover luminosity.

There are two obvious ways to obtain the Coma/Virgo distance ratio
through the GCLF data:

\noindent (a) compare the GCLF turnovers
of just the two central cD galaxies, M87 and NGC 4874; or

\noindent (b) compare the mean GCLF turnovers 
of all Virgo giant ellipticals with
the mean for our two Coma ellipticals.

\noindent The latter route
has the advantage of greater statistical weight over many galaxies, 
while the former has the
advantage of matching similar types of galaxies as strictly as possible,
albeit at the cost of greater internal uncertainty.
As we will see below, the two approaches turn out to agree quite well.

\subsection{M87 vs. NGC 4874}

At various times in the literature, concerns have
been raised regarding possible dependences of the GCLF shape and peak on
environment, galaxy type, and cluster metallicity.
By restricting the comparison to M87 and NGC 4874 alone, 
we can minimize these worries.  
Both are centrally placed giant ellipticals with cD envelopes, and though
they are not identical in size or luminosity (NGC 4874 is a magnitude
more luminous, and Coma is distinctly richer than Virgo), the comparison
is certainly closer than with an average E galaxy.
M87 has one of the best-studied of GCS luminosity functions:  
repeated observations to ever-increasing depth and radial coverage
have extended the GCLF far past the turnover and have now made it
possible (for example) to define GCLFs separately for the two parts 
of its clear bimodal color distribution or as a function of
galactocentric distance.

For M87, the best determination of the GCLF turnover 
is by Kundu \etal\ (1999)\nocite{kun99}, who 
find $V^0$(M87) $= 23.67 \pm 0.06$
from their deep WFPC2 photometry, after subtraction of an adopted
foreground extinction $A_V = 0.067 \pm 0.04$ mag.  
Subtracting $V^0$(M87) from our NGC 4874
determination, we immediately derive $\Delta\mu_0$(Coma-Virgo) $=4.15
\pm 0.14$ (with the plausible assumption that both galaxies are at or
near the physical centers of their clusters).

A further sophistication on this comparison can be taken if we look more
narrowly at the metallicity and radial dependences.
Several deep photometric studies 
(\cite{gri86}; \cite{lau86}; \cite{mcl94}; \cite{har98b}; 
\cite{kun99}) show radial
trends in the GCLF turnover can be ignored as the level of variation is
at the $\pm 0.2-$mag level or less.  For the metallicity issue,
Whitmore \etal\ (1995)\nocite{whi95} and Kundu \etal\ (1999) \nocite{kun99} 
find that the peak of the
M87 GCLF has a {\it small} -- and perhaps not significant --
dependence on color, where the GCLF may be $\sim 0.1$ mag fainter
for the redder, more metal-rich half of the bimodal color distribution.
However, NGC 4874 differs from M87 in that its globular cluster color
distribution is {\it not} bimodal:  our two-color data 
indicate (Figure~\ref{fig:VI}) that the mean
color of the NGC 4874 GCS is very much like that of the bluer, more metal-poor
half of the M87 cluster system. 
For the metal-poor component, Whitmore \etal\ 
find $V^0$(M87, blue) $ \simeq 23.6\pm0.1$ where the uncertainty
includes both the photometric and fitting uncertainty combined as
random errors.\footnote{Whitmore \etal\ do not state an
uncertainty for their determination of $V^O$; we take it
to be their quoted fitting error for the entire GCLF,
increased by $\sqrt{2}$ to allow for
the smaller sample size of the metal-poor cluster population relative to
the total population.}
In a new reduction of the same images, Kundu \etal\ \nocite{kun99}
find entirely similar results.  
If we then take $V^0$(M87, blue) $= 23.6\pm0.1$ and subtract it from
the NGC 4874 turnover, we obtain $\Delta\mu_0$(Coma-Virgo) $= 4.22 \pm
0.16$.  However, it is not obvious that the metallicity offset is real
or that it should be applied.  Two counterarguments are the following:

\begin{figure*}
\epsscale{0.5}
\plotone{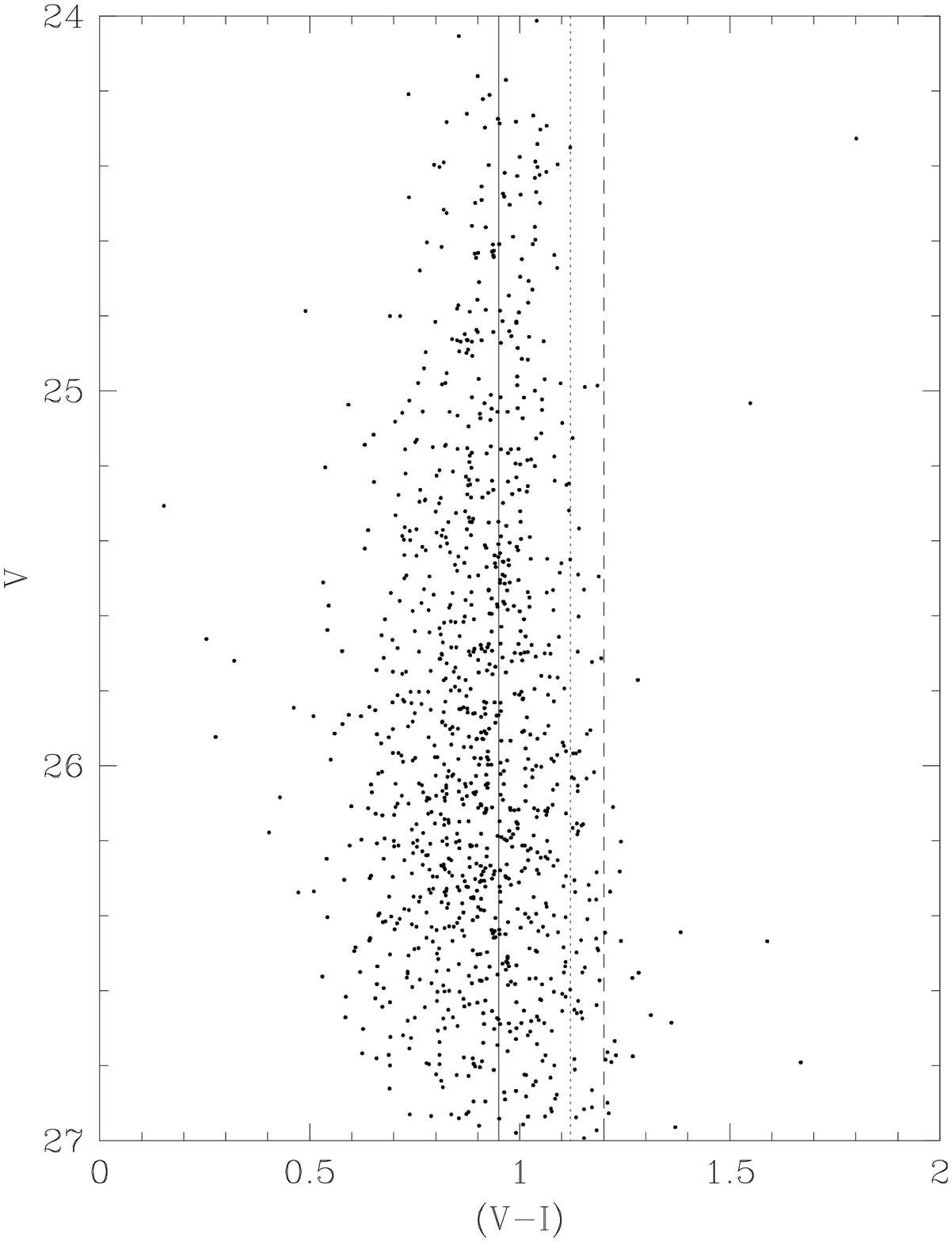}\caption[f5.eps]{$(V-I)$ colors of starlike objects within 70'' of the center 
of NGC 4874.  Objects with photometric measurement uncertainties less
than $\epsilon(V-I) = 0.07$ are plotted.
The mean color of the globular cluster population is $<(V-I)> \simeq 0.95$, 
which is similar to the metal-poor cluster population of M87 (solid line). 
For comparison, the long dashed line is the mean color of the metal-rich
component of the M87 GCS and the dotted line is the mean $(V-I)$ of the entire
M87 GCS (Whitmore \etal\ 1995; Kundu \etal\ 1999).
\label{fig:VI}}
\end{figure*}

\noindent (a) The Virgo and Fornax ellipticals display a wide variety
of globular cluster color distributions -- unimodal, bimodal, red, or blue
(e.g., \cite{nei99}; \cite{ajh94}; \cite{geb99}; \cite{kis97}).
These obvious
differences in the metallicity distributions, however, do not result
in any clearly correlated changes in the GCLF turnover points within
a scatter of $\pm 0.15$ mag (see below).

\noindent (b) The color distribution for the clusters in IC 4051, the
other Coma giant, is entirely unimodal and {\it red}, just the opposite
of NGC 4874 (\cite{bau97}; \cite{woo99}).  
Yet its GCLF turnover differs from NGC 4874 by only
0.1 mag and is, if anything, {\it brighter}, contrary to the case for M87.

\begin{figure*}
\epsscale{0.5}
\plotone{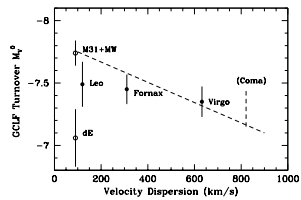}
\caption[f5i.eps]{GCLF turnover luminosity $M^0_V$ plotted
against the velocity dispersion of the host cluster of galaxies.  
{\it Filled symbols} represent the giant E galaxies in Leo, Fornax, and
Virgo, from the most recent determinations of their distances and
GCLF observations. {\it Open symbols} show the disk and dwarf E members of the
Local Group.  The location of the two Coma galaxies is shown by
the vertical dashed line at right, while the proposed correlation from
Blakeslee \& Tonry (1996) is shown as the long dashed line.
See text for discussion.  \label{fig:disp}}
\end{figure*}

An alternate argument put forward by Blakeslee \& Tonry (1996) \nocite{bla96} 
is that the turnover of the GCLF depends on environment, represented
by the velocity dispersion of the group of galaxies it is in. 
They suggest that the turnover becomes {\it less} luminous within {\it
more} massive galaxy clusters (that is, ones with higher velocity
dispersion); and if so, then we would be overestimating the true Coma
turnover luminosity since it is a more massive environment than Virgo.

The empirical correlation for this effect
is shown in Figure~\ref{fig:disp} (a replotting of Blakeslee \& Tonry's
Figure~2, with the more recent values for $M_V^0$ for 
a variety of GCLFs).   The Leo data are taken from \cite{kw99},
and M31, the Milky Way, and the Local Group dwarf ellipticals
from the compilation of 
\cite{har99}.  Rather than mixing all types of parent galaxies,
we restrict our attention to only the giant
ellipticals (in the large disk galaxies and dE's, the globular
clusters may indeed have experienced very different tidal fields
and thus degrees of dynamical destruction).
We find a very much more modest trend than did Blakeslee \& Tonry,
well within the $\sim 0.2-$mag scatter expected from the data.
Furthermore, it is not obvious just where along the horizontal axis
of this graph Coma ``belongs'', since there is abundant evidence
(see Paper II for a summary) that it has assembled from a variety of
smaller original clusters, structural traces of which are visible today in the
galaxy spatial distributions and intracluster gas.

Our provisional conclusion is that, although {\it slight}
dependences of GCLF turnover on such factors as metallicity, radius,
and environment may be expected on current
theoretical grounds, the weight of direct
observational evidence indicates no such trends above the measurement
scatter of the method.  
For the {\em present}, we therefore choose not to apply
any such adjustments to the GCLFs until such time as they are
considerably more well established on strict empirical grounds.

\subsection{The Virgo Ellipticals vs. the Coma Ellipticals}

Our second approach
is simply to take the average turnover luminosity of all
the ellipticals in both clusters.  In doing so, we deliberately average
over (or in effect, ignore) any differences in metallicity, elliptical type,
or sample differences in radial location.  It is {\it empirically} true
that all these effects are minor:  for a sample of 17 giant ellipticals
with well measured GCLFs and fundamental distance calibrations (Virgo, 
Fornax, and a handful of others at similar distances), the turnover
luminosity exhibits a galaxy-to-galaxy
scatter of $\pm 0.15$ mag (\cite{har99};
\cite{whi96}).  For six Virgo gE's including M87, Table \ref{tab:vlist}
shows $V^0 = 23.71\pm0.03$, a number which is indistinguishable 
from our adopted value for the M87 peak by itself though it is internally
more precise.  Thus, by this approach we obtain
$\Delta \mu$(Coma-Virgo) $= (28.77 \pm 0.10) - (23.71 \pm 0.03) = 4.06
\pm 0.11$.  

All the results derived here for the relative Coma/Virgo modulus agree
with each other within their internal uncertainties, and
there is no evidence that any of the factors mentioned above
(cD vs.~normal giant elliptical; cluster metallicity; radial location)
produce measurable systematic biasses to within the $\pm0.1-$mag fitting
uncertainties built into the GCLF method.  We therefore adopt the last
of the determinations listed above ($\Delta\mu = 4.06 \pm 0.11$), which
has the highest internal precision and greatest statistical security of
sample size.  Adding this to our adopted Virgo distance modulus, we obtain
$\mu$(Coma) $= 35.05 \pm 0.12$ or $d = (102 \pm 6)$ Mpc (where 
the quoted error represents only
the internal uncertainty of the method).

Earlier literature
(see, e.g., \cite{cap90}; \cite{san90}; \cite{van92} for reviews of
a range of methods) tended to give $\Delta \mu$(Coma $-$ Virgo)
in the range $3.7 - 3.8$.  However, 
our determination of 4.06 is similar to other, more
recent determinations that rely directly on the giant ellipticals:  
for example, SBF analysis of ellipticals in Coma
and Leo I (\cite{tho97}) and the fundamental plane comparison for the
Leo and Coma ellipticals (\cite{hjo97}) give $\Delta \mu$(Coma $-$ Leo)
$\simeq 4.8 - 4.9$.  Subtracting $\Delta \mu$(Virgo $-$ Leo) $ = 0.87 \pm 0.1$
(\cite{fer99a}) then gives $\Delta \mu \simeq 4.0$ to 4.1 for the
Virgo-to-Coma step, in agreement with our findings.  

The GCLF turnover luminosity that we implicitly adopt along with
this distance calibration is, {\it from the Virgo cluster alone},
$M_V^0 = -7.26 \pm 0.06$ (internal uncertainty).  For Fornax alone,
our turnover luminosity would be $M_V^0 = -7.45 \pm 0.06$.  The true
external uncertainties on both of these estimates, particularly
for Fornax, are likely to be near $\pm0.2$ mag.

\section{The Hubble Constant}

We can now estimate $H_0$.
The redshift of Coma corrected for Local Group peculiar motion
(e.g., \cite{col96})
is $cz = 7100$ km s$^{-1}$, with
a likely uncertainty of $\pm 200$ km s$^{-1}$ taking into account the
degree to which the peculiar velocities of the Milky Way and Local
Supercluster relative to the CMB are known.  We note that this mean
Coma velocity explicitly excludes the outlying NGC 4839 subgroup,
which would have biased the mean to slightly higher levels (see
\cite{col96}).

From Hubble's law, $cz = H_0 \cdot d$, we obtain
$$H_0 = cz \cdot 10^{(5-0.2 \mu_0)} \, {\rm km} \, {\rm s}^{-1} \, {\rm Mpc}^{-1} \, . $$
Corrections for the geometric curvature parameter $q_0$ 
are negligible (if $q_0$ is in the range $\sim 0.0 - 0.5$, at the Coma
redshift of $z = 0.0237$ the resulting uncertainty in $H_0$ is only 0.3
percent).  Putting in our distance modulus and redshift for Coma, we obtain 
$H_0 = 69$ km s$^{-1}$ Mpc$^{-1}$.

The net uncertainty in our result must include not only the 
internal uncertainties of the Coma/Virgo relative distance and the 
CMB-corrected Coma velocity, but also 
a series of other external uncertainties.  The true (external) error
is dominated by the uncertainty in the fundamental distance scale
that enters our discussion through the Virgo distance.  
This calibration depends in turn on a multitude of connected
issues such as the distance to the LMC and the parallaxes of nearby
standard candles (RR Lyraes, Cepheids, subdwarfs).  A full discussion of
these issues is far beyond the scope of this paper, but our reading of the
literature suggests that the external uncertainty in the Virgo distance
modulus may now be at the level of $\pm 0.2$ mag.  

Table~\ref{tab:err} contains a complete review of our error budget.  
Combining the uncertainties in quadrature, and without distinguishing
them rather arbitrarily as ``internal'' or ``external'',
we derive for our final estimate
$$H_0 = (69 \pm 9) \, {\rm km} \, {\rm s}^{-1} \, {\rm Mpc}^{-1} \, . $$
\begin{deluxetable}{lll}
\tablewidth{0pt}
\tablecaption{\label{tab:err} Error budget for our $H_0$
measurement}
\tablehead{\colhead{...}  & \colhead{uncertainty source} & \colhead{km s$^{-1}$ Mpc$^{-1}$}}

\startdata

$\pm 0.25$ $\Delta q_0$ & curvature parameter $q_0$: & $ \pm 0.2 $\\

$\pm 0.04$ mag & foreground extinction: & $ \pm 1.3$  \\

$\pm0.05$ mag & WFPC2 photometric zeropoint: & $ \pm 1.6$ \\

$\pm 200$ km s$^{-1}$ & Coma CMB-corrected redshift:& $\pm 2.0$ \\

$\pm 0.1$ mag & turnover luminosity (blue vs.~red): & $\pm 3.3$ \\

$\pm 0.11$ mag & GCLF turnovers (Coma {\it minus} Virgo): & $\pm 3.6$ \\

$\pm 0.2$ mag & Virgo distance modulus: & $\pm 6.8$ \\
\enddata
\end{deluxetable}

The Coma galaxies are easily the most remote ones for which
GCLF turnover levels have been directly measured, and thus they exert the
most leverage on any estimates of $H_0$ with this technique.  However,
as a last useful check on the result, we can display the GCLF results
for several galaxies and clusters in a classic ``Hubble diagram'':  that is, a
plot of redshift against the apparent magnitude of the turnover.
Hubble's law can be rewritten as
$$ {\rm log} \, cz = 0.2 V^0 + {\rm log} H_0 - 0.2 M_V^0 - 5 $$
where $M_V^0$ is the GCLF turnover luminosity for giant E galaxies,
and $H_0$ is expressed in its usual units.

\begin{figure*}
\epsscale{0.5}
\plotone{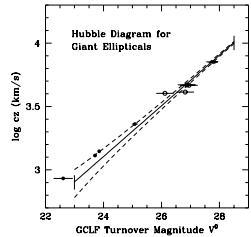}
\caption[f6.eps]{Hubble diagram for the GCLF technique.  Here, galaxy redshift
$cz$ (km s$^{-1}$) is plotted against the apparent magnitude $V^0$ of
the GCLF turnover (corrected for foreground extinction $A_V$).
Data points are taken from Table~\ref{tab:groups}.
The solid line has a slope 0.2 (see Section 5 of the text) and a
zeropoint corresponding to $H_0 = 70$ and $M_V^0 = -7.3$.
Vertical tick marks at either end of the line show the range
$\Delta H_0 = \pm 9$ allowed by our determination.
Filled circles are the galaxies with direct GCLF observations past
the turnover point, while open circles are the four galaxies with 
deduced $V^0$ values from SBF analysis.  The {\it dashed lines} show the
expected range within which points would fall if the redshifts $cz$ are
uncertain by $\pm 200$ km s$^{-1}$; see text. 
\label{fig:groups}}
\end{figure*}

Relevant data for a total of 10 galaxies or groups are listed in
Table \ref{tab:groups} and plotted in Figure \ref{fig:groups}\begin{deluxetable}{lllll}
\tablewidth{0pt}
\tablecaption{GCLF Turnover Levels in Distant Ellipticals}
\tablehead{\colhead{Cluster} & \colhead{Galaxy} & \colhead{Redshift $cz$}
& \colhead{$V^0$(GCLF)} & \colhead{Sources}\tablenotemark{a} }
\startdata
Leo I & 2 gE's & $850$ km s$^{-1}$ & $22.61 \pm 0.33$ & 1 \\
Virgo & 6 gE's & $1300$ & $23.71 \pm 0.03$ & 2 \\
Fornax & 6 gE's & $1400$ & $23.85 \pm 0.04$ & 2 \\
NGC 5846 & NGC 5846 & $2300$ & $25.08 \pm 0.10$ & 3 \\
Coma & IC 4051 & $7100$ & $27.75 \pm 0.20$ & 4 \\
Coma & NGC 4874 & $7100$ & $27.82 \pm 0.12$ & 2 \\
A 262 & NGC 705 & $4650$ & $26.95 \pm 0.3$  & 5 \\
A 3560 & NGC 5193 & $4020$ & $26.12 \pm 0.3$  & 5 \\
A 3565 & IC 4296 & $4110$ & $26.82 \pm 0.3$  & 5 \\
A 3742 & NGC 7014 & $4680$ & $26.87 \pm 0.3$  & 5 \\
\tablenotetext{a}{Sources:  (1) Harris 1990
(2) This paper
(3) Forbes \etal\ 1996a
(4) Woodworth \& Harris 1999
(5) Lauer \etal\ 1998 }
\label{tab:groups}
\enddata
\end{deluxetable}

(See Harris 1999 for the first use of this Hubble diagram for GCLFs).
In this table, the entries for Virgo and Fornax are the weighted
mean $\langle V^0 \rangle$ values listed previously, while the entry
for the nearby Leo I system is the average of NGC 3377
and 3379 from \cite{kw99}.  The cosmological recession velocities 
($cz$) assume a Local
Group infall to Virgo of $250 \pm 100$ km s$^{-1}$ (e.g., \cite{ford96};
\cite{ham96}; \cite{jer93}, among others).
The mean radial velocities of each galaxy or group are taken from
Faber \etal\ (1989)\nocite{fab89}, Girardi \etal\ 
(1993)\nocite{gir93}, Huchra (1988)\nocite{huc88}, 
Binggeli \etal\ (1993)\nocite{bin93}, Hamuy \etal\ (1996)\nocite{ham96}, 
and Colless \& Dunn (1996)\nocite{col96}.

The last four entries in Table \ref{tab:groups} are from \cite{lau98}
for brightest cluster ellipticals (BCG's) in four Abell clusters.
These authors assume
a Gaussian GCLF shape with $\sigma_V = 1.4$ (strictly comparable with
what we use here), and then derive the $V^0$ which best fits the
measured luminosity function.  

The solid line in Figure \ref{fig:groups} is the solution for $H_0 = 70$
and $M_V^0 = -7.3$ (the mean luminosity of the GCLF turnover with our
adopted Virgo calibration).  The raw scatter of the points around the
line is $\pm 0.25$ mag rms, an independent estimate of
the accuracy of the GCLF technique that is encouragingly similar to 
our global error budget summed above.  Interestingly, the four nearest
points (Leo I, Virgo, Fornax, and NGC 5846) are all on the high side of
the line; with our adopted $cz$ values as they stand, 
these four points by themselves would give $H_0 \simeq 78$.  
Conversely, they would all lie precisely on
the mean line for $H_0 = 70$ if their CMB-frame redshifts were reduced
by 200 km s$^{-1}$ from the values listed in Table \ref{tab:groups}. 
The dashed lines in Figure
\ref{fig:groups} show the range that points would be expected to fall
within purely by random velocity errors of $\pm 200$ km s$^{-1}$.
The importance of obtaining galaxies at the largest possible distances
to circumvent systematic errors on $H_0$ is evident.

\section{Summary}

\begin{enumerate}
\item Using the Hubble Space Telescope WFPC2 we have obtained deep
images of the GCS of NGC 4874, the central cD in COMA.
Photometry with DAOPHOT/ALLFRAME provides us with a measurement of
the globular cluster luminosity function which reaches a 50\%
completeness at apparent magnitude $V = 28.2$.

\item Using both a constrained $\chi^2$ and maximum likelihood
analysis we find that the best-fit Gaussian for the GCLF of NGC 4874
has a turnover at $V^0= 27.82\pm0.13$ for a Gaussian of
width $\sigma_V=1.4$, and corrected for foreground absorption and
$K-$correction.  Combining this measurement with previous 
determinations of the
GCLF turnover for another Coma elliptical, IC 4051, yields an 
average turnover value of $V^0 = 27.77\pm0.10$ for Coma.

\item Comparing the Coma turnover value with the average value for
six well studied ellipticals in Virgo implies a relative 
Coma/Virgo distance modulus $\Delta\mu_0 = 4.06 \pm 0.11$.
Adding the Virgo distance as calibrated from the mean of the SBF, PNLF, TRGB
and Cepheid techniques results in an intrinsic Coma distance modulus of
$\mu_0 = 35.05\pm0.13$ (internal error).

\item Adopting a recession velocity of $v_r = 7100 \pm 200$ km s$^{-1}$
for Coma yields
a Hubble constant of $H_0 = 69$ km s$^{-1}$ Mpc$^{-1}$.
A full error budget suggests that the true uncertainty in this estimate
is $\pm 9$ km s$^{-1}$ Mpc$^{-1}$ or 13\%; the dominant term is the
uncertainty in the distance to Virgo, which we use as our fundamental
calibrating point for GCLFs in giant ellipticals.
 
\end{enumerate}

Further improvements in the range of allowed values for $H_0$ will come with
(a) the addition of more galaxies in Coma and other distant clusters,
(b) continued effort to determine the true dependence of the GCLF peak 
on metallicity, and (c) more certain calibrations of the distances to
the LMC, Fornax, Virgo, as the most fundamental stepping stones in
the distance ladder.

\acknowledgements

This research was supported through grants from
the Natural Sciences and Engineering Research Council of Canada.

{}

\end{document}